\documentclass{article}
\usepackage{graphicx}
\begin{document}
\twocolumn
[
\begin{center}
{ \Large \bf
A simple analytical model for dark matter halo structure and adiabatic contraction

\medskip}
{ \large \bf \sl
E.A.\,Vasiliev

\medskip}
{\sl
Lebedev Physical Institute, Moscow, Russia\\
e-mail: eugvas@lpi.ru}

\begin{quote} \small
A simple analytical model for describing inner parts of dark matter halo is considered. 
It is assumed that dark matter density is power-law. The model deals with dark matter distribution function
in phase space of adiabatic invariants (radial action and angular momentum). Two variants are considered for the 
angular part of the distribution function: narrow and broad distribution.
The model allows to describe explicitly the process of adiabatic contraction of halo due to change of gravitational 
potential caused by condensation of baryonic matter in the centre.
The modification of dark matter density in the centre is calculated, and is it shown that the standard algorithm 
of adiabatic contraction calculation overestimates the compressed halo density, especially in the case of strong
radial anisotropy. 
\end{quote}
\end{center}
%              darkmat.  MWmass    MWhalo
{\small PACS: 95.35.+d, 98.35.Ce, 98.35.Gi\\[0.5cm]

}

]{

\section{\large\sc Introduction}

The structure and evolution of dark matter halos is widely discussed in the recent years (e.g. \cite{BM})
A large number of papers has been devoted to the investigation of spatial density profiles of dark matter halos,
$\rho(r)$.
both numerical \cite{Navarro, Moore05} and analytical \cite{LMKW}. 
Recently several papers appeared that described general properties of solutions of Jeans equation 
\cite{DM,AWBBD}.
The Jeans equation for a sperically symmetric self-gravitating system reads as follows:
\begin{equation}  \label{Jeans}
\frac{d}{dr} (\rho\,{\sigma_r}^2) + \frac{2\beta}{r} (\rho\,{\sigma_r}^2) + \rho\frac{G\,M(r)}{r^2} = 0 \,.
\end{equation}
Here ${\sigma_r}^2$ and ${\sigma_t}^2$ are the radial and tangential velocity dispersion, 
$\beta = 1-\frac{{\sigma_t}^2}{2\,{\sigma_r}^2}$ is the Binney's anisotropy parameter (systems with 
$0<\beta<1$ are radially anisotropic, with $\beta<0$ -- tangentially anisotropic, and $\beta=0$ is the isotropic case).

An additional assumption is the following empirical property of dark matter halos: 
the generalized phase-space-{\it like} density of dark matter particles in a halo is a power-law in radius:
\begin{equation}  \label{gpsd}
\rho(r)/{\sigma_r}^\epsilon(r) \propto r^{-\alpha}  \,,
\end{equation}
which is established both in N-body calculations \cite{TN} and some analytical models \cite{AWBBD}.

For a closed set of equations one also needs to specify $\beta$. Here two different approaches may be adopted:
in \cite{AWBBD} isotropy was assumed ($\beta=0$), and in \cite{DM} the following relationship 
between $\beta$ and the logarithmic slope of density profile $\gamma = d \ln\rho(r)/d r$ was taken:
\begin{equation}  \label{betagammarelation}
\beta = \beta_0 + b(\gamma-\gamma_0)  \,.
\end{equation}
This kind of relation has been proposed in \cite{HM} based on empirical analysis of different numerically 
treated situations.

These investigations demonstrated that under these assumptions there exist only few possibilities for a self-consistent 
halo model. The most realistic is the set of models with density profiles having inner and outer power-law 
asymptotes (so-called Zhao $\alpha\beta\gamma$ models \cite{Zhao}).

However, this approach, being very fruitful, does not describe explicitly the distribution function. 
It operates only with its second moment with respect to velocity, i.e., the velocity dispersion. 
The most complete description of halo structure can be made in terms of phase-space distribution function $f(r,v)$.
One attempt to deal with full velocity distribution function (VDF) was made in \cite{Tsallis}, where an Eddington 
inversion formula was applied to obtain VDF for power-law density profiles. However, this study was limited to 
isotropic case.

An approach based on certain physical assumptions regarding the form of the distribution function seems more promising.
It avoids postulating empirical assumptions like (\ref{gpsd}, \ref{betagammarelation}), which, nevertheless, still
hold in many cases. 
It is convenient to express the distribution function in terms of adiabatic invariants during slow evolution of the 
gravitational potential. This allows to describe easily the process of adiabatic contraction of dark matter halo caused
by condensation of baryonic matter in its centre, which occurs during the formation of a galaxy.
This approach is adopted in the present paper.

The paper is organized as follows. In the second section we describe two models of inner halo structure and explain
physical arguments for the underlying assumptions concerning the dark matter distribution function. 
In the third section we calculate the velocity anisotropy coefficient. Furthermore, in the fourth section we consider 
the halo response to the adiabatic compression and compare it with other studies. 
Finally, the conclusions are presented.

\section{\large\sc The phase-space structure of model halo}

Dark matter halos are extended complex objects which formed in different physical processes (growth of initial 
perturbations, collapse, hierarchical merging, and later -- baryonic contraction). 
We are intererested in the central area of a halo, which presumably is less influenced by merging and more -- by 
condensation of baryons.
Most of existing dark halo models assume power-law behaviour of density profile in the centre of a halo. 
\begin{equation}  \label{rho}
\rho_d(r) = K\,r^{-\gamma}\,, \;  1\le\gamma<2 \,.
\end{equation}
We shall adopt this assumption, having in mind that our analysis applies only to the inner area of a halo, interior to
the radius where the logarithmic density slope starts to deviate from the constant value $\gamma$. 
The anisotropy coefficient is also taken to be constant in this area ($0\le \beta <1$, 
according to analytical calculations and numerical simulations \cite{LM}).
Under these assumptions, $f$ depends on $E$ also in a power-law form \cite{BT}.

We write distribution function in terms of action-angle variables. The action variables are the angular momentum $m$,
its projection onto $z$ axis $m_z$, and radial action $I_r$, defined as $I_r = \frac{1}{2\pi}\oint v_r(r)\,dr$.
The distribution function is quasi-stationary and hence does not depend on angle variables, and due to spherical 
symmetry, neither it depends on $m_z$. The radial action in power-law potential is well approximated by the following 
expression:
\begin{equation}  \label{I_r}
I_r = A\,E^{\frac{2-\gamma/2}{2-\gamma}}\,\left(1-\frac{m}{m_c(E)}\right) \,,
\end{equation}
where $m_c(E)$ is the angular momentum of circular orbit with given energy \cite{GS}.

We consider two variants of inner halo structure, which differ in the dependence of distribution function on angular 
momentum.

In the model A the angular momentum $m$ of a particle is proportional its radial action:
\begin{equation}  \label{m_vs_I}
m=l_0\,I_r  \,.
\end{equation}
This relation, according to \cite{GZ}, emerges as a result of dynamical mixing in the centre of a collapsing dark
matter halo. Initial density peak, which gives birth to the gravitationally bound object, is taken to be triaxial 
ellipsoid. Gravitational collapse and subsequent phase mixing, as shown in this paper, lead to formation of 
spherically symmetric structure, in  which particles have angular momentum directly linked to their initial distance
from centre, and, hence, with their radial action.
The quantity $l_0$ in expr. (\ref{m_vs_I}) is related to orbital eccentricity of particles: $l_0=0$ corresponds 
to pure radial orbits, and $l_0 \to \infty$ -- to circular.

The distribution function in model A is expressed as follows:
\begin{equation}  \label{DFa}
f(I_r,m) = f_0\,E^{1/2}\,\delta(m^2-{l_0}^2{I_r}^2) \,.%,\quad \lambda = \frac{2-\gamma}{4-\gamma}\,.
\end{equation}

In the model B the distribution function is given by expression
\newcommand{\Beta}{\mathrm B}
\begin{equation}  \label{DFb}
f(E,m) = f_0\,E^\mu\,m^{-2\Beta}\,,\quad \mu=\frac 1 2 -\frac{4-\gamma}{2-\gamma}(1-\Beta)\,.
\end{equation}
This is the simplest generalization of isotropic model for the case of arbitrary radial anisotropy \cite{BT, EA}.

Therefore, the two models considered represent, in some sense, two limiting cases -- narrow and broad distribution
by angular momentum for fixed energy.
In what follows, we express particle energy through variables $I_r$ and $m$ according to (\ref{I_r}), 
and rewrite the distribution function in action-angle variables.

Notice that we do not consider possible change of distribution function in the subsequent evolution, hierarchical 
clustering and merging, which is not significant for the inner parts of a halo, as shown in analytical calculations
\cite{Dehnen} and numerical simulations \cite{KZK}.

Due to scale invariance, the expression (\ref{gpsd}) is essentially satisfied, with
$\alpha=(1-\frac{\gamma}{2})\epsilon$. For $\gamma<2$ and $\epsilon>2$ 
the power-law index $\alpha$ is slightly greater than 2. This is similar to ESIM (Extended Secondary Infall models)
described in \cite{AWBBD}, in which the density has also power-law form at small radii.

\section{\large\sc Velocity anisotropy}

As long as the distribution function is given in action-angle variables, one can compute velocity distribution function 
at each radius, and, consequently, obtain radial and tangential velocity dispersion values ${\sigma_r}^2$ and 
${\sigma_t}^2$. Obviously, they should have also power-law form $\sigma^2 \propto r^{2-\gamma}$. 

In the model A the anisotropy coefficient $\beta$ is related to the density slope $\gamma$ via the additional 
parameter $l_0$ (see Fig.1). This relation can be approximated as
\begin{equation}  \label{betagamma}
\beta \approx \gamma-1 - 0.27\,l_0\,\gamma \,.
\end{equation}
Hence, for a given $l_0$ the relation between $\gamma$ and $\beta$ has linear form (\ref{betagammarelation}), 
though the coefficients differ from \cite{HM}.
The upper bound on $\beta$ equals $\gamma-1$, in agreement with \cite{Hansen}.

In the model B the anisotropy coefficient $\beta$ is identical to $\Beta$ in expr. (\ref{DFb}) \cite{BT}. 
Notice that it is possible to have $\beta > \gamma-1$, if we recall that power-law density profile (\ref{rho})
breaks down at large radii.
In this case the velocity dispersion is determined by density profile cutoff scale, and depends on radius as
$\sigma^2 \propto r^{\gamma-2\beta}$. Upper bound on $\beta$ is then $\gamma/2$, according to \cite{EA}. 
In the model A it is not possible, since it follows from expr. (\ref{I_r}, \ref{m_vs_I}) that the particle velocity
at each radius lies in a limited range, proportional to $\sqrt{\Phi(r)}$, and hence,
for sufficiently small radii the cutoff scale does not determine the velocity dispersion.

\section{\large\sc Response to adiabatic contraction}

It is well-known that central parts of galaxies are dominated by baryonic matter. The response of dark matter halo to 
the baryonic infall usually is calculated using the adiabatic approximation. 
The common algorithm for computing compression is that of Blumenthal et al. \cite{Blum}, in which dark matter 
particle's angular momentum is conserved. It is strictly applicable only to particles on circular orbits. 
In this method, the final radius $r_f$ of a dark matter particle is related to its initial radius $r_i$ by 
the following expression:
\begin{equation}  \label{blumm}
(M_{fin,DM}(r_f)+M_{fin,bar}(r_f))\,r_f = M_{in,DM}(r_i)\,r_i \;,
\end{equation}
where $M_{fin,DM} + M_{fin,bar}$ is the total mass of dark matter and baryons after contraction, and $M_{in,DM}$ is 
the initial dark matter mass inside a given radius.

\begin{figure}[t]
$$\includegraphics{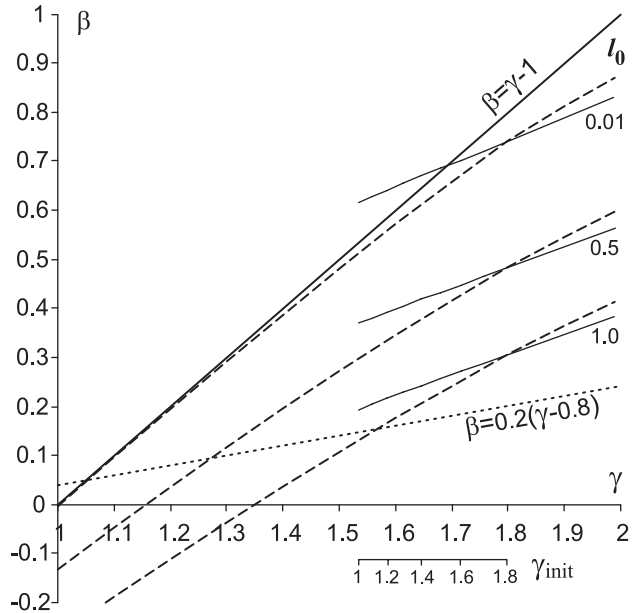} $$
\caption{\footnotesize 
Anisotropy $\beta$ vs. density slope $\gamma$ relation in model A: \protect\\
dashed lines -- initial anisotropy for different values of $l_0$; \protect\\
solid lines -- anisotropy after compression. Corresponding lines for the same $l_0$ cross at $\gamma=\Gamma=1.8$.
Horizontal axis is the slope $\gamma$ for original (uncompressed) halo and $\gamma'$ for compressed, which depends on 
$\gamma$ via (\ref{gammaprime}). Corresponding original values of $\gamma$ are shown in the bottom auxiliary axis.
\protect\\
Thick line -- upper limit $\beta=\gamma-1$; \protect\\
dotted line -- Hansen \& Moore relation\cite{HM}}
\end{figure}
 
In general case of non-circular orbits, the radial action is also conserved, being an adiabatic invariant. 
However, usually this does not help much, as the radial action is not known in general case. 
%In this case, one may substitute $r_{max}$, the orbit apocenter, for $r_i$ and $r_f$ in (\ref{blumm}). 
An improved method proposed in \cite{GKKN} involves the combination $M(\bar r)\,r$, where $\bar r$ is the 
orbit-averaged radius of a particle. This yields less compressed halos, in better agreement with numerical 
simulations. Alternatively, an iterative method can be used for computing density profile and radial action 
in baryon-induced potential \cite{SM}.

In our simple models we are able to explicitly consider the adiabatic contraction, since we have our distribution 
function already written in terms of adiabatic invariants (provided that the characteristic timescale of the 
compression is much greater than dynamical timescale).
All we need is to calculate radial action in new potential and recompute density profile. 
In many cases we may assume the baryonic density profile to be also power-law in radius:
$\rho_b(r) = K_B\, r^{-\Gamma}$, with index $\Gamma \le 2$ (close to isothermal profile). 
This again applies only to central area of a galaxy, the bulge. We assume $\Gamma > \gamma$.

\begin{figure}[t]
$$\includegraphics{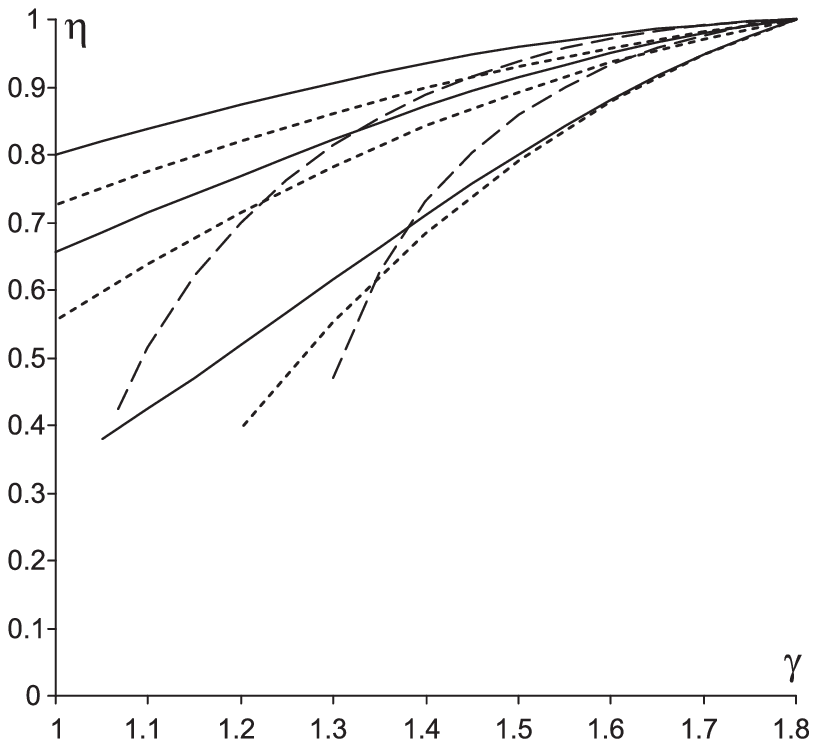} $$
\caption{\footnotesize $\eta$ -- the ratio of our model's density normalization to that of the standard 
adiabatic contraction model, for different values of anisotropy coefficient $\beta$. \protect\\
Dashed lines -- model A, $\beta=0$ and 0.25 (from top to bottom); \protect\\
solid and dotted lines -- models A and B correspondingly, for $\beta=0$, 0.25 and 0.5 (top to bottom), 
in model A the anisotropy coefficient $\beta'$ after contraction is used.}
\end{figure}

It is quite obvious that the modified dark matter density profile is also power-law: $\rho_d'(r)=K' \,r^{-\gamma'}$.
The new power-law index $\gamma'$ is expressed as follows:
\begin{equation}  \label{gammaprime}
\gamma' = \frac{3\Gamma+\gamma-\Gamma\gamma}{4-\gamma} \,.
\end{equation}

The power-law index is essentially the same as in Blumenthal's method, but the normalization coefficient is different. 
We introduce the quantity $\eta = K'/K_{std}'$, the ratio of our model's normalization to that of Blumenthal's method 
(see Fig.2). It appears that $\eta < 1$, in agreement with the results of \cite{GKKN} and \cite{SM}, 
where it was found that halos with radial motion are compressed less than predicted by Blumenthal et al. algorithm.
The quantity $\eta$ decreases with decrease in $\gamma$ and increase in $\beta$. Numerical simulations in \cite{CLMW} 
also confirm this result; these authors explain it by the fact that for fixed $\beta$ a greater value of $\gamma$ 
increases the dominance of circular orbits over radial ones, and this makes Blumenthal's method more applicable.
(Note that when $\beta$ tends to $\gamma-1$ in model A or to $\gamma/2$ in model B, the density normalization is 
more and more affected by density cutoff radius, so that values of $\eta$ are underestimated).

For quantitative calculations we take Milky Way bulge and dark halo as an example. The bulge density profile and 
normalization are taken from \cite{CS} and are the following: 
$\rho_b=0.6\cdot 10^9\, M_\odot/{\rm kpc}^3 (r/1\,{\rm kpc})^{-\Gamma}$, $\Gamma=1.8$.
The dark matter halo is normalized to have density 0.3 GeV/cm$^2$ at r=8.5 kpc \cite{Merritt}.
(Note that it yields different initial mass of dark matter inside 1 kpc, ranging from $4\cdot 10^8\;M_\odot$ for 
$\gamma=1$ to $3\cdot 10^9\;M_\odot$ for $\gamma=1.8$. The baryonic mass within 1 kpc is $6\cdot 10^9\;M_\odot$).

We calculate the ratio of new to old dark matter mass within radius 1 kpc and plot it in Fig.3. Again we see that with
increasing $\beta$ (more radial orbits), the actual compression ratio becomes smaller. 
The greater the difference between $\gamma$ and $\Gamma$, the more is the increase in dark matter mass, i.e.
the compression is more noticeable for halos with lower power-law index of initial density cusp. 

\begin{figure}[t]
$$\includegraphics{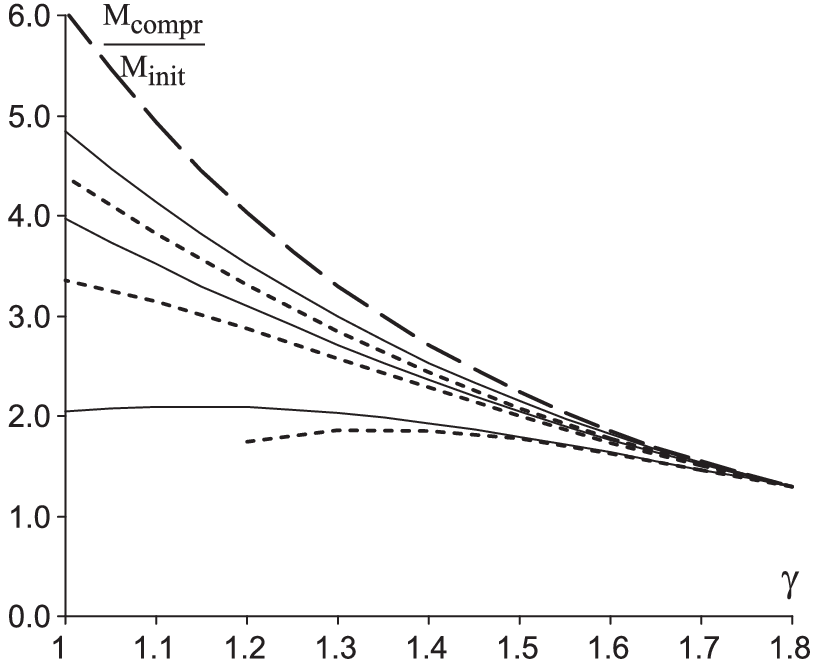} $$
\caption{\footnotesize Ratio of compressed to initial dark matter mass within 1 kpc:\protect\\
long dashed -- compressed by standard (Blumenthal) algorithm;\protect\\
solid and dotted lines -- models A and B correspondingly, for $\beta=0, 0.25$ and 0.5 (from top to bottom); 
in model A the anisotropy coefficient $\beta'$ after contraction is used.}
\end{figure}

And finally, we calculate the velocity anisotropy parameter $\beta'$ for modified halo (see Fig.1). 
In model A it appears to be generally greater than for corresponding uncompressed halo, meaning that velocity 
becomes even more radially biased. 
At the same time, the relation between $\beta'$ and $\gamma'$ becomes more flat (the same effect can be seen from Fig.2 
of \cite{HM} when comparing pure cosmological simulations and dark matter with cooling baryons simulations).

Having in mind that setting $\Gamma=\gamma=\gamma'$ effectively does not change profile, we approximate velocity
an\-iso\-tro\-py parameter as
\begin{equation}
\beta' \approx \beta(\Gamma) + 0.4(\gamma'-\Gamma) \,.
\end{equation}

In model B the anisotropy coefficient slightly decreases (at most by 0.04 for $\gamma=1$, $\beta=0$).
We note that due to increase in $\beta'$ for model A it resembles more the model B in the reaction for adiabatic 
compression, if we plot the compression coefficient as a function of $\beta'$, not $\beta$ (solid and dashed lines in 
Figs.2 and 3).

\section{\large\sc Conclusion}

We have considered two simple analytical models for structure of central part of dark matter halo. 
In both models, the distribution function depends on energy in a power-law form, and the dependence on angular momentum 
is $\delta$-function in model A and power-law in model B.
The simplicity of the models allows us to calculate some interesting properties of the halo, namely: the velocity 
anisotropy and response to adiabatic contraction due to baryonic infall. 
These calculations help to get qualitative insight into halo properties and lead to the following conclusions:
\begin{enumerate}
\item The more radially biased is the velocity, the less compressed is the halo. 
\item The shallower is initial density profile (less value of $\gamma$), the less is the degree of compression 
compared to that of Blumenthal's method. 
Standard model of adiabatic contraction systematically overestimates the effect of contraction. 
\item The model predicts a moderate enhancement (2 to 4 times) of dark matter mass in the bulge. 
The shallower is the initial profile, the greater the increase of dark matter mass.
This result significantly increases the estimates of possible annihilation radiation flux from the center 
of the Galaxy \cite{GS, Merritt, ZV}.
\end{enumerate}

I am grateful to Maxim Zelnikov for helpful discussion on the topic of the paper,
to Steen Hansen for pointing out some interesting issues and to anonymous referee for valuable comments and proposals. 
This work was supported by Landau Foundation (For\-schungs\-zent\-rum J\"ulich) and Russian Fund for Basic 
Research (project nos. 01-02-17829, 03-02-06745).

}
\end{document}